\pdfoutput=1
\documentclass[sigconf]{acmart}

\AtBeginDocument{%
  }

\copyrightyear{2026}
\acmYear{2026}
\setcopyright{cc}
\setcctype{by}
\acmConference[CCS '26]{Proceedings of the 2026 ACM SIGSAC Conference on Computer and Communications Security}{November 15--19, 2026}{The Hague, Netherlands}
\acmBooktitle{Proceedings of the 2026 ACM SIGSAC Conference on Computer and Communications Security (CCS '26), November 15--19, 2026, The Hague, Netherlands}
\acmDOI{10.1145/3830454.3846631}
\acmISBN{979-8-4007-2871-6/2026/11}

\usepackage{enumitem}

\graphicspath{{./graphics/}}

\usepackage{xspace}

\newcommand{\rpmcroot}[0]{\textit{Write Root Key Register}\xspace}
\newcommand{\rpmcupdate}[0]{\textit{Update HMAC Key Register}\xspace}
\newcommand{\rpmcget}[0]{\textit{Request Monotonic Counter}\xspace}
\newcommand{\rpmcinc}[0]{\textit{Increment Monotonic Counter}\xspace}
\newcommand{\rpmcstatus}[0]{\textit{Read Extended Status}\xspace}

\newcommand{\rpmcincshort}[0]{\textit{Increment}\xspace}

\newcommand{\code}[1]{\texttt{#1}}

\begin{document}

\renewcommand{\sectionautorefname}{Section}
\renewcommand{\subsectionautorefname}{\sectionautorefname}
\renewcommand{\subsubsectionautorefname}{\sectionautorefname}
 
\title{Hello Again: Exploiting Replay Protected Monotonic Counters to Brute-Force Windows Hello PINs}
\titlenote{This is a slightly updated version of the paper published at ACM CCS~'26 (\url{https://doi.org/10.1145/3830454.3846631}). The section \hyperref[sec:changelog]{\emph{Changes from the CCS '26 Version}} lists all changes.}

\author{Christian Werling}
\affiliation{%
  \institution{Technische Universität Berlin}
  \city{Berlin}
  \country{Germany}
}
\email{cwerling@sect.tu-berlin.de}

\author{Matti Finder}
\affiliation{%
  \institution{Technische Universität Berlin}
  \city{Berlin}
  \country{Germany}
}
\email{mfinder@sect.tu-berlin.de}

\author{Hans Niklas Jacob}
\affiliation{%
  \institution{Technische Universität Berlin}
  \city{Berlin}
  \country{Germany}
}
\email{hnj@sect.tu-berlin.de}

\author{Julian Prein}
\affiliation{%
  \institution{Technische Universität Berlin}
  \city{Berlin}
  \country{Germany}
}
\email{j.prein@tu-berlin.de}

\author{Niclas Kühnapfel}
\affiliation{%
  \institution{Technische Universität Berlin}
  \city{Berlin}
  \country{Germany}
}
\email{kuehnapfel@tu-berlin.de}

\author{Jean-Pierre Seifert}
\affiliation{%
  \institution{Technische Universität Berlin}
  \city{Berlin}
  \country{Germany}
}
\email{jean-pierre.seifert@tu-berlin.de}

\begin{abstract}
Windows Hello is the primary authentication mechanism on over one billion Windows 11 devices, supporting PINs as short as four digits. Such low-entropy PINs are viable only because the TPM's anti-hammering protection mitigates brute-force attacks. Most modern x86 platforms rely on firmware TPMs (fTPMs) that persist their non-volatile (NV) state, including the lockout state, to SPI flash. On newer platforms, rollback prevention of this state is delegated to the flash chip via Replay Protected Monotonic Counters (RPMCs), a JEDEC-standardized mechanism operating over a physically accessible bus.

We present the first practical security analysis of RPMC in shipping x86 systems. Using low-cost interposer hardware, we identify multiple implementation flaws, two of which fundamentally undermine RPMC's security model: First, Winbond SPI flash -- the leading NOR flash vendor with over 25\% market share -- leaks its HMAC key under voltage glitching, enabling arbitrary counter forgery. Second, AMD's NV controller does not verify RPMC increments, so an interposer can suppress them unnoticed. Moreover, AMD platforms without RPMC flash appear to lack NV rollback protection altogether: on two AMD desktops, reflashing an old image resets the fTPM's lockout counters.

Both flaws enable rollbacks of the fTPM's NV state (AMD, Intel, or Pluton), including its lockout counters. We demonstrate this end-to-end on locked, encrypted hardware: our setup finds a 4-digit Windows Hello PIN in 40 hours instead of 69 days. A compromised PIN also grants access to device-bound keys and Hello-based enterprise SSO. The same primitive defeats BitLocker's TPM+PIN pre-boot protector, reducing the worst-case time for a default-length 6-digit PIN from 19 years to 20 days.
\end{abstract}



\maketitle

\section{Introduction}
\label{sec:introduction}

While user authentication on PCs traditionally revolved around passwords, smartphones have commoditized biometrics and short numeric PINs to authenticate a variety of operations, from unlocking the screen to auto-filling passwords to issuing payments.
In order to protect low-entropy PINs from the brute-force attacks of the earlier smartphone generations \cite{defcon_into_the_droid, mdsec_ios_bruteforce,elcomsoft_ios_bruteforce}, Apple, Google, and Samsung have built dedicated hardware and multi-layered software systems around PIN authentication \cite{apple_platform_security,pixel_6_titan_security,samsung_knox_security_whitepaper,android_gatekeeper}.
They all employ escalating timeouts after failed verification attempts, some even allowing permanent lockouts after a number of incorrect PINs, and the option to wipe sensitive data altogether.
Despite such measures, previous research has shown that even standardized hardware-based implementations are not immune to protocol or hardware vulnerabilities \cite{wd_rpmb_protocol_vulns_2020,fukami_keyless_2024,rpmb_exploits_nfi_2024}.

Windows Hello aims to close this security and usability gap on PCs and offers PINs as short as 4 numeric digits as the primary authentication method or as a fallback for biometrics.
Along with \textit{Windows Hello for Business}, Hello is advertised as the central authentication method, beyond unlocking the laptop, that also authenticates the use of other device-bound credentials and serves as a Single Sign-On (SSO) provider for enterprises.
Microsoft needs to secure these PINs equally, but across a much broader set of hardware than smartphone vendors do.
Hello documentation claims that the (in Windows 11 mandated) TPM 2.0 makes Windows Hello PINs ``anti-hammering'' protected and hardware-bound, thwarting brute-force and phishing attacks, respectively \cite{whfb_overview}.
\looseness=-1
Previous research on Hello has already shown that its protection without a TPM is inadequate \cite{kim_security_2021,elcomsoft_notpm_nosecurity,haddad_attacking_2023,gerontakis_vulnerability_2025}.
Therefore, Hello relies on PIN-authorized TPM sealing to protect the user's private key material used for the above-listed authentication features.
TPM 2.0's ``Dictionary Attack Protection'' \cite[16.8]{tpm2spec_architecture} is the critical technology to, by default, employ a 32 try limit across reboots that re-enables one try every 10 minutes \cite{microsoft_tpm_fundamentals}.\footnote{On two of our test systems (Framework and Acer, both Intel Core Ultra 5), we instead observed a recovery time of two hours (see \emph{Lockout Configurations} in \autoref{sec:bruteforce_lockout_conf}).}

While premium laptops have traditionally opted for dTPMs in the past, possibly due to better on-paper security or regulatory requirements \cite{nsa_tpm_use_2024}, dTPMs suffer from an exposed bus allowing passive attackers to sniff sensitive data \cite{andzakovic_bitlocker_2019,pulsesec_bitlocker_2020} and active attackers to mount reset attacks \cite{kauer_oslo_2007, werling_dtpm_2025}.
A more popular type of TPM implementation now is firmware TPMs (fTPMs).
These run in Trusted Execution Environments (TEEs), either on the CPU's security coprocessors (AMD PSP, Intel CSME), or in an isolated execution layer of the main core (Qualcomm QTEE/QSEE on TrustZone), making their communication with the x86 client application much harder for attackers to inspect.
In 2024, Microsoft adopted this trend and collaborated with SoC vendors to integrate a custom coprocessor, Microsoft Pluton, that is envisioned to act as a key storage provider for Windows, but can also act as an fTPM \cite{microsoft_pro_understanding_pluton}.
Microsoft requires it on all Copilot+-certified laptops \cite{microsoft_pluton, microsoft_introducing_copilotplus_2024}.
However, in July 2026, Microsoft's documentation silently added a phase-out notice to Pluton's fTPM functionality on AMD and Qualcomm \cite{microsoft_pluton_as_tpm}, falling back to the vendor's firmware TPM (as was already the case on Intel) or a discrete TPM.
Previous work has shown that fTPMs may be compromised via side channels \cite{tpmfail_2020} or through a compromise of their host security processor \cite{faultpm_2023}.
However, we identified another threat vector:
To persist their state across reboots, fTPMs store it along with other non-volatile (NV) data on SPI flash.
This data is encrypted and integrity protected via device-unique keys used by AMD's and Intel's NV controllers \cite{faultpm_2023, csme_whitepaper}.
While the data, therefore, cannot be inspected or altered by an attacker with access to SPI flash under normal circumstances, they could still save the ciphertexts and replay them.
Microsoft seems well aware that platform vendors like AMD and Intel must secure their NV controllers from such rollback attacks by explicitly requiring rollback protection in both their Windows 10 and Windows 11 hardware compatibility programs \cite{win10_hcp, win11_hcp}.
\looseness=-1
Nonetheless, we have not found any practical research on the security of these.

\paragraph{Replay Protected Monotonic Counters}

SPI flash has evolved over the years to deliver NOR flash content at ever-increasing speeds.
NV controllers loading firmware code and data usually employ code-signing mechanisms to provide secure boot: loading and executing only trusted software.
``Replay Protected Monotonic Counter'' (RPMC) is an additional JEDEC specification that extends SPI flash with security counters.
Our research shows that systems with RPMC use it to secure fTPM and other non-volatile storage of AMD's, Intel's, and Microsoft's own fTPM.

\subsection{Contributions}

This paper presents the first public security analysis of both RPMC controllers and flash chips deployed in shipping x86 systems.
Assuming physical access, we evaluate attacks on the SPI bus used for RPMC communication, both towards the SPI flash and the SPI controller.

\begin{itemize}
  \item First security evaluation of three RPMC flash vendors: Winbond, Macronix, GigaDevice (\autoref{sec:flash_vulnerabilities}):
    \begin{itemize}
      \item \emph{Integer Overflow}: On Winbond, Macronix, and GigaDevice, an increment to the maximum integer value leads to overflows of the counters, contradicting RPMC's monotonicity (minor impact).
      \item \emph{Counter Corruption}: On GigaDevice flash, voltage glitching allows manipulation of returned RPMC counter values (minor impact).
      \item \emph{HMAC Key Leak}: On Winbond flash, voltage glitching leaks valid RPMC signatures and, more critically, the authentication keys (\textbf{major impact}, see below).
    \end{itemize}
  \item First security evaluation of the dominant x86 vendors' NV controllers (Intel, AMD) with and without RPMC (\autoref{sec:interposer-attacks}):
    \begin{itemize}
      \item \emph{Naive Rollback}: All tested Intel NV controllers and all RPMC-backed AMD NV controllers withstand naive rollbacks of the flash contents (including fTPM state); two AMD systems (Zen~2 and Zen~5) without RPMC-capable flash do not (\textbf{major impact}).
      \item \emph{HMAC Key Reuse}: Intel's RPMC-based NV controllers reuse the same \textit{Key Data} across all counters and sessions, weakening counter authentication (minor impact).
      \item \emph{Unverified Increment}: \looseness=-1 AMD's RPMC-based NV controllers do not verify increment success (\textbf{major impact}, see below).
    \end{itemize}
  \item NV-state rollback, rendering fTPM Dictionary Attack Protection ineffective, confirmed end-to-end against live targets via two independent attack paths using a custom, low-cost SPI hardware interposer (\autoref{sec:nv-state-rollback-attack}): \emph{AMD's Unverified Increment} on an HP OmniBook 3 and \emph{Winbond's HMAC Key Leak} on a Framework Laptop 13.
  \item End-to-end attack against Windows Hello: PIN brute-forcing demonstrated on an AMD laptop at around 4.14 tries/minute, resulting in a worst-case time of 40 hours for a 4-digit numerical PIN (\autoref{sec:brute-forcing-with-nv-rollbacks}).
  \item End-to-end attack against BitLocker: PIN brute-forcing demonstrated on an Intel laptop via the Winbond \emph{HMAC Key Leak} rollback path at around 34.1 tries/minute, resulting in a worst-case time of 20 days for a default-length 6-digit PIN (\autoref{sec:bitlocker-bruteforce}).
\end{itemize}

Note: The two attack paths above are independent of the target application; we pair them with Windows Hello and BitLocker purely as two illustrative end-to-end proofs of concept.

\paragraph{Disclosure}
We informed all directly and indirectly affected parties of these issues, i.e., Winbond, Macronix, GigaDevice, JEDEC, Intel, AMD, and Microsoft. Unfortunately, none of them has committed to fixing the root-cause vulnerabilities; only Microsoft announced a mitigating policy change, raising the recommended minimum Hello PIN length to 6 digits.
Their detailed statements and remediation status are discussed in \autoref{sec:discussion}.

\section{Background}

\subsection{Trusted Platform Modules}
\label{sec:bg_tpm}

Trusted Platform Modules, as defined by the TPM 2.0 Library Specification, are a standard for cryptographic modules providing a variety of security features \cite{tpm2spec_intro}.
We will shortly cover its three most important features for this research:

\subsubsection{Sealed Storage}
TPM \emph{sealed storage} allows the TPM to encrypt and integrity-protect an arbitrary data blob.
This data can only be unsealed by the same TPM that created it and only when a set of release conditions is satisfied~\cite{tpm2spec_architecture}.
  
\subsubsection{Policy Sessions}
Policies are a TPM authorization mechanism that implements, e.g., the release conditions for TPM sealed storage.
Each policy is represented by a cryptographic digest computed offline as the cumulative hash of a sequence of policy assertions.
Software interfacing with the TPM always does so in the context of a \emph{policy session} also represented by one or more policy digests.
If one of these digests matches the policy digest of, e.g., a sealed storage, it serves as a valid authorization to unseal said storage.
Policy digests in a policy session can be constructed by, e.g., entering a passphrase or measuring system state, or combined out of other policy digests.
This allows policies to implement arbitrarily complex, cryptographically bound authorization policies~\cite{tpm2spec_architecture}.

\subsubsection{Dictionary Attack Protection}
\label{sec:bg_anti_hammering}
\label{sec:tpm_dap}
\label{sec:bg_tpm_dictionary_attack}

The TPM enforces a hardware-level \emph{dictionary attack (DA) protection} mechanism to limit exhaustive searches against authorization values (passphrases or PINs) stored on the device~\cite{tpm2spec_architecture}.
The lockout logic is governed by four 32-bit NV-persistent variables counting failed authorization attempts \cite{tpm2spec_architecture}.
These can be configured in parameters such as their \texttt{maxTries} and \texttt{recoveryTime}, which determine after how many failed authorization attempts a \textit{Lockout Mode} is entered and after what time a failed attempt is forgotten \cite{tpm2spec_architecture}.
Setting the \texttt{recoveryTime} to a large value, e.g. $2^{32}-1$ seconds, allows entering an effectively permanent lockout.

\subsubsection{Firmware TPMs}
\label{sec:bg_ftpms}

While it used to be common that TPMs were realized in dedicated chips connected to the SoC via an LPC or SPI bus (discrete TPMs or dTPMs), the industry now has largely shifted towards potentially more cost-effective firmware  TPMs (fTPMs) running in security coprocessors.
Moreover, despite AMD and Intel offering fTPMs by default in their SoCs, Microsoft started collaborating with SoC vendors to integrate a Microsoft-designed security core named Microsoft Pluton \cite{microsoft_pluton}.
dTPMs suffer from a central vulnerability: their exposed bus towards the SoC.
In that regard, fTPMs do not trivially expose their communication to the outside as their communication happens inside the SoC.
However, in contrast to dTPMs, AMD's, Intel's, and Microsoft Pluton's fTPMs do not have dedicated on-chip memory securely storing their non-volatile state.
Instead, they rely on the SoC to provide them with non-volatile (NV) storage.
Previous work on the AMD PSP and Intel CSME analyzed internals of how NV storage is implemented in these security coprocessors \cite{faultpm_2023, sklyarov_mfs_explained}:
On both platforms, the NV storage (including fTPM state) stores its data alongside the UEFI firmware on the SPI flash.
To protect it from manipulation by an attacker with SPI access, its contents get encrypted and integrity protected using device-unique keys.
Additionally, when the NV controller boots up, it must ensure that the NV state read from flash is the most recent one.
Otherwise, features such as the fTPM Dictionary Attack Protection, which heavily rely on consistent state progression across reboots, are vulnerable to rollback attacks.

\subsection{Replay Protected Monotonic Counters}
\label{sec:rpmc}\label{sec:rpmc_chronicle}
RPMC is a specification of JEDEC for SPI flash devices that builds on top of the SPI bus protocol  \cite{jesd260}.
RPMC serves as an additional building block in securing the contents of said devices.
Devices supporting this standard are extended with commands enabling the usage of multiple monotonic counters \cite{jesd260}.
It is closely related to one part of another JEDEC specification: the Replay Protected Memory Block (RPMB) of eMMC \cite{jesd84}.

Integration of RPMC has been an ongoing effort.
The groundwork was a document by Intel in 2013 \cite{intel_flash_hardening_2013}; this document was later copied in 2021 to form the official JESD260 specification \cite{jesd260}, making it the de facto standard way to extend flash chips with security counters. 
In recent times, support for these counters has been added to major open source projects like Coreboot\footnote{https://review.coreboot.org/c/coreboot/+/84837}, but Tesla's fork\footnote{https://github.com/teslamotors/coreboot/tree/tesla-4.12-amd} of it contained artifacts hinting at its usage for even longer.
By now, RPMC is used in commercial off-the-shelf laptops to secure the NV state of TPMs and protect sensitive EFI variables (e.g., Secure Boot enabled) against rollback attacks.
In combination with traditional code signing techniques, RPMC can also be used to prevent firmware rollback, but we were unable to find real-world examples as of now.

\begin{figure}[ht]
    \centering
    \includegraphics[width=0.92\linewidth]{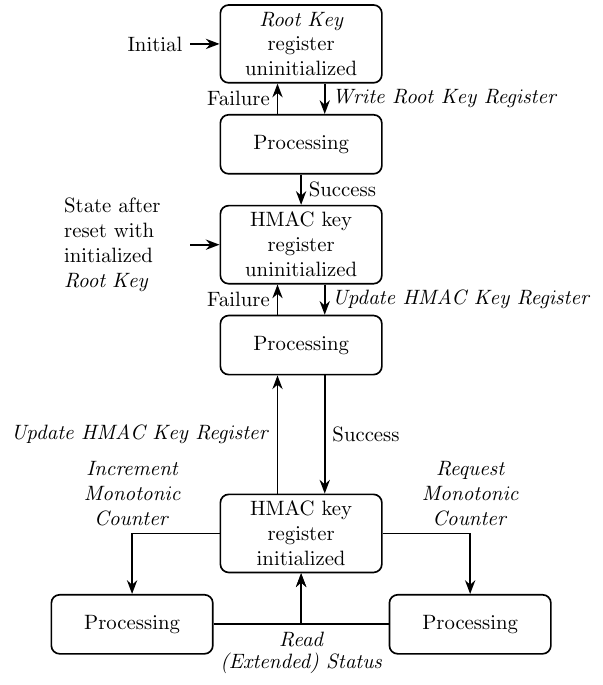}

    \caption{Per Counter RPMC Command Flow}
    \Description[A graph showing the flow of RPMC commands]{}
    \label{fig:rpmc_command_flow}
\end{figure}

RPMC consists of five main commands; the regular command flow can be seen in \autoref{fig:rpmc_command_flow}.
To prove their authenticity, commands are appended with a Message Authentication Code, also called a signature within the specification, generated using HMAC with SHA256 as the underlying hash function \cite{jesd260}.
RPMC uses two types of keys: the \emph{Root Key} serves as the persistent pre-shared secret that authenticates users of a counter, and the HMAC Key serves as a temporary session key for an already authenticated party.
Each counter's \emph{Root Key} can only be initialized once.\footnote{The exception is a \emph{Root Key} of all \texttt{0xff}, which can still be overwritten later \cite{jesd260}. The counter value remains at its previous value in this case.}
After \emph{Root Key} initialization, these counters start in the ``HMAC key register uninitialized'' state \cite{jesd260}.

\paragraph{\rpmcroot}
This command is used to initialize the \emph{Root Key} of a specific counter.
If successful, the selected counter is set to 0, and the \emph{Root Key} is stored in the \emph{Root Key} register of that counter \cite{jesd260}.

\paragraph{\rpmcupdate}
\label{sec:rpmc_cmd_update_hmac_key}
This command only works if the root-key register for the specific counter was populated.
It is used to transmit 4 bytes of \textit{Key Data}, which form the HMAC Key (HMAC of the \textit{Key Data} using the \emph{Root Key}), for the current session \cite{jesd260}.
This HMAC Key is used by the controller to authenticate any further RPMC commands (\rpmcinc, \rpmcget,  and also the \rpmcupdate itself) and by the chip itself to sign the data returned by \rpmcstatus  \cite{jesd260}.

\paragraph{\rpmcinc}
\label{sec:rpmc_cmd_increment}
A successful execution of this command increments the selected counter by 1 \cite{jesd260}.
A valid request requires including the current value of the selected counter to ensure that the request, and consequently, the signature, is always different on subsequent calls of this command.

\paragraph{\rpmcget}
\label{sec:rpmc_cmd_get}
This command is used to prepare the retrieval of a selected counter value \cite{jesd260}.
Each request includes a randomization Tag to prevent replay attacks.
The Tag can be populated with any 12 bytes of data and will be echoed during the retrieval with \rpmcstatus.

\paragraph{\rpmcstatus}
\label{sec:rpmc_cmd_read_rpmc_status}
This command can be issued after previous commands to retrieve progress information or the requested counter value.
The response signals the execution state of any previous RPMC command, indicating whether the chip is still busy, has successfully completed the command, or if an error (e.g., signature mismatch) occurred.
When \rpmcget was issued beforehand, and the response indicates a success, the response includes three additional fields: Tag, Counter Value, and Signature.

\subsubsection{Attacks \& Threat Model}
\label{sec:rpmc_threat_model}
In the specification, JEDEC mentions, ``The current definition of the Serial Flash does not offer any protection against HW attacks'' \cite{jesd260}.
JESD260 itself echoes this by requiring that increment operations be performed in a power glitch aware manner \cite{jesd260}, a requirement Winbond's datasheets reproduce almost verbatim \cite{winbond_w25r128jv}.
As RPMC is introduced to fix this gap, the main threat model for these chips are attackers that already have physical access to a device containing such a chip. 

Previous security research exists that targets a similar use of counters for RPMB, highlighting protocol weaknesses \cite{wd_rpmb_protocol_vulns_2020}, exploitation with Chip-off techniques \cite{rpmb_exploits_nfi_2024}, and glitching possibilities using EMFI \cite{fukami_keyless_2024}.

\subsection{Windows Security}

\subsubsection{Disk Encryption}
\label{sec:bg_disk_encryption}

Full disk encryption is a well-established method to protect data at rest using industry-grade encryption; however, its security mainly depends on the encryption key used.
Microsoft BitLocker is Windows' built-in disk encryption software, allowing a transparent encryption of the main volume.
An interplay of firmware-based Secure Boot (\emph{UEFI Secure Boot}) and a TPM enables automatic decryption when firmware and the Windows bootloader are in known-good states, providing data security without additional friction for the user \cite{bitlocker_countermeasures}.

\paragraph{TPM+PIN}

For higher security needs, BitLocker offers a \emph{TPM+PIN} protector: in addition to the verification of the known-good states, the Windows Boot Manager prompts for a PIN that authorizes the TPM to \emph{Unseal} the volume master key.
This PIN is subject to the same Dictionary Attack Protection counters as Windows Hello (\autoref{sec:bg_anti_hammering}), but, unlike Hello, without any additional Windows-imposed try or timeout limit layered on top.
On our test systems, the startup PIN had to be at least 6 digits; however, an administrator can lower this to a 4-digit floor by explicitly enabling and configuring the \emph{Configure minimum PIN length for startup} policy \cite{microsoft_bitlocker_faq}.

\subsubsection{Windows Hello}
\label{sec:windows_hello}\label{sec:whfb_sso}

Once Windows has booted up, its lockscreen protects the -- now unencrypted -- Windows environment.
Windows Hello is a more recent security feature by Microsoft introduced in Windows 10 and the only login type for online accounts on Windows 11 \cite{winhello2hashcat_banaanhangwagen}.
It manages the user's access to ``their identity provider, applications, and services'' \cite{windows_hello_ess}, including the use as a FIDO2 security key.
In contrast to using a password to unlock the system and authenticate sensitive operations, Hello allows users to use a fingerprint or face recognition service for authentication \cite{windows_hello_ess}.
\emph{Windows Hello for Business} (WHFB) extends Windows Hello, which is local to the machine, with external identity providers \cite{bsi_whfb_dissected_2026}. It ``provides enterprise-grade security and management capabilities, including device attestation, certificate-based authentication, and conditional access policies''.
Notably, WHFB ``is considered two-factor authentication'' by Microsoft, signaling that no additional factors are needed \cite{whfb_overview}.

With or without biometrics, Hello asks users to set up a numeric or alphanumeric PIN at least 4 digits long.
While such a PIN may have much lower entropy than a traditional password, Microsoft claims that Windows Hello prevents brute-force attacks through the use of a TPM and by employing escalating timeouts \cite{microsoft_tpm_fundamentals}.
In fact, ``anti-hammering protection'', as Microsoft calls it, is active on two levels:
The Windows lockscreen also employs a 5-try per boot limit before asking the user to reboot the system.
Then, the device enters a lockout state that escalates with each subsequent failure cycle \cite{whfb_faq}.
Microsoft claims that the Hello PIN entry is additionally protected by the TPM's Dictionary Attack Protection (\autoref{sec:tpm_dap}).
This configures the TPM to lock after 32 authorization failures and to forget one authorization failure every 10 minutes \cite{microsoft_tpm_fundamentals}.
According to our research, Microsoft does not allow permanent lockouts after a defined number of tries, neither for Windows Hello nor WHFB.
However, businesses can adapt their PIN policies to require stronger PINs than the default 4-digit numeric PIN Microsoft allows \cite{whfb_policy_settings}.

Previous research on Windows Hello has shown that on devices without a TPM, Windows Hello's authentication data is not properly protected \cite{kim_security_2021, elcomsoft_notpm_nosecurity, haddad_attacking_2023}.
While some research on WHFB sheds light on the inner workings of the authentication, others have developed specialized tools to assess PIN security over the network \cite{{gerontakis_vulnerability_2025}}.
A qualitative study on Windows Hello in a small business found that while participants perceived it as more usable and responsive than traditional Windows authentication, they tended to prefer PINs over biometrics due to limited hardware support or being out of reach of the laptop \cite{farke_exploring_2022}.

\paragraph{Hardware Requirements}
\label{sec:windows_hcp_rpmc}

Windows 11 famously made the presence of a TPM mandatory \cite{win11_hcp}.
However, TPMs were already the object of Windows 10's Hardware Compatibility Program (HCP) \cite{win10_hcp}.
It leaves open whether a discrete or firmware TPM must be available.
However, some relevant changes to the document have happened to its underlying security assumptions since the oldest available version from 2015:
While the first two versions (1511, 1607) do not require RPMC, version 1703 made RPMC for rollback prevention mandatory:
\begin{quote}
RPMC for non-discrete TPMs (consumer 16 MB parts) is a requirement for 2019.
RPMC for UEFI (commercial 32 MB parts) is a requirement for 2020.
\end{quote}
Versions 1709, 1803, and 1809 kept this phrasing until version 1903 (i.e., the version of March 2019) generalized this requirement shortly before it became effective:
\begin{quote}
\textbf{Effective October 1, 2020} This requirement is Mandatory for platforms that support specific hardware  capabilities for replay protection. Confidential \& replay-protected storage: If a TPM uses external memory for non-volatile storage of TPM state (including seeds, proof values, \& dictionary attack variables), movement of the TPM state to and from the NV memory should include protections of that data to ensure confidentiality and integrity of the data and to mitigate against rollback attacks. This is generally accomplished by encrypting the data, applying a Message Authentication Code, and storing the resulting record in replay-protected storage \textbf{such as Replay Protected Memory Block (RPMB) or Replay Protected Monotonic Counter (RPMC)}.
\end{quote}

While Windows 11's HCP also leaves open which rollback prevention method is used, it makes it mandatory for specific platforms only \cite{win11_hcp}:
\begin{enumerate}
      \item Intel platforms with Intel PTT firmware version 600.* or higher.
      \item AMD platforms with Zen~2 (4000 series) or newer.
      \item Qualcomm platforms with RPMB-capable storage.
\end{enumerate}

Regarding Microsoft's own fTPM, Microsoft Pluton, the document specifies that certain AMD platforms must support a TPM 2.0 device, and that the supported TPM 2.0 devices are discrete TPM 2.0 and Pluton-based firmware TPM 2.0.
Notably, the corresponding Intel System Guard section instead lists ``discrete TPM 2.0 or integrated TPM 2.0 (Intel PTT)'' -- Pluton is not called out on the Intel side.
Note that in July 2026, Microsoft added a phase-out notice for Pluton's fTPM functionality on AMD and Qualcomm platforms, falling back to the vendor's firmware TPM or a discrete TPM \cite{microsoft_pluton_as_tpm}.

\section{Flash Attacks}
\label{sec:flash_vulnerabilities}

\begin{table*}[t]
    \centering
    \caption{Overview of Tested Memory Chips with RPMC Support}
    \label{tab:tested_chips}
    \begin{tabular}{c c c c c c c}
        \toprule
        Manufacturer & Series & Package Identifier & Supply Voltage & Integer Overflow & HMAC Key Leak & Counter Corruption\\
        \midrule
        Winbond & W25R128JV & SIQ & 3 V & \checkmark & \checkmark & \scalebox{0.85}[1]{$\times$} \\
        Winbond & W25R128JW & PIQ & 1.8 V & \checkmark & \checkmark & \scalebox{0.85}[1]{$\times$} \\
        \midrule
        Macronix & MX77L & 12850FM2I40 & 3 V & \checkmark & \scalebox{0.85}[1]{$\times$} & \scalebox{0.85}[1]{$\times$} \\
        Macronix & MX77U & 12851FM2I42 & 1.8 V & \checkmark & \scalebox{0.85}[1]{$\times$} & \scalebox{0.85}[1]{$\times$} \\
        \midrule
        GigaDevice & GD25LR256 & FYIGE & 1.8 V & \checkmark  & \scalebox{0.85}[1]{$\times$} & \checkmark\\
        \bottomrule
    \end{tabular}
\end{table*}

In this section, we will present our security findings when targeting SPI flash chips with RPMC support in isolation.
Later sections test the RPMC controller and flash in integration.
We analyzed flash chips across different operating voltages from three major manufacturers: Winbond, GigaDevice, and Macronix.
The chips tested are listed in \autoref{tab:tested_chips}.

\paragraph{Attacker Capabilities}

Since RPMC was designed to offer protection against hardware attacks \cite{jesd260}, we assume an attacker with the following capabilities:
\begin{itemize}
    \item Ability to attach a controller, e.g., flash programmer, to the chip and send arbitrary commands as well as read their responses
    \item Ability to influence the chip's supply voltage, allowing for the insertion of voltage glitches
    \item Ability to interact with the flash chip for a prolonged amount of time
\end{itemize}

\paragraph{Attack Setup}
\label{sec:flash_attack_setup}
\begin{figure}[h]
    \centering
    \includegraphics[width=0.9\linewidth]{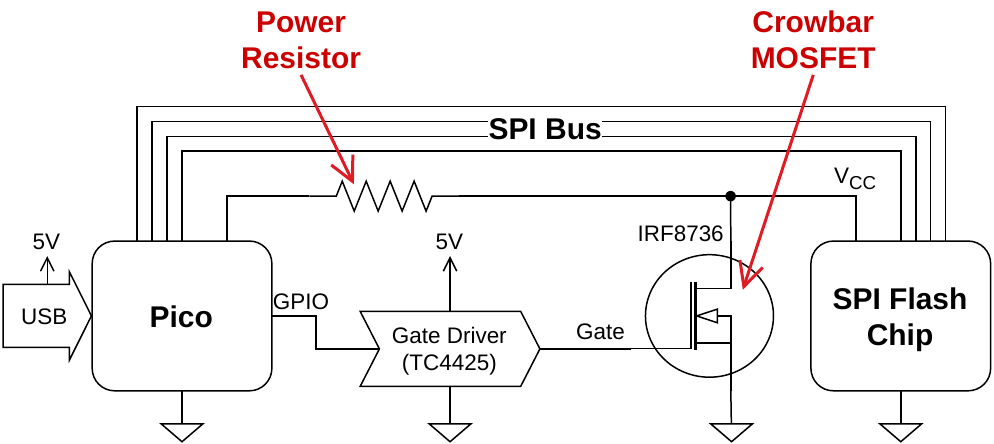}
    \caption{Schematic Diagram of the Attack Setup}
    \Description[A circuit schematic showing a crowbar voltage-glitching setup between a Raspberry Pi Pico and an SPI flash chip]{A Raspberry Pi Pico controls a gate driver that switches a MOSFET (the crowbar) wired across the SPI flash chip's power supply through a power resistor, allowing the Pico to pull the flash chip's supply voltage down while both share the same SPI bus.}
    \label{fig:glitch_setup}
\end{figure}

We started testing with a breadboard setup, shown in \autoref{fig:glitch_setup}.
The setup revolves around a Raspberry Pi Pico 2W microcontroller acting as the SPI controller, the SPI flash chip (the device under test), an IRF8736 power MOSFET, and an appropriate gate driver to optionally meddle with the SPI flash's supply voltage.
The Pico is running with the flag \textit{PICO\_USE\_FASTEST\_SUPPORTED\_\allowbreak CLOCK} set to run at 150 MHz (as of April 2026); changing this flag affects the parameters we highlight later.

The setup forms a crowbar circuit on the flash chip's $V_{cc}$, enabling it to quickly and precisely pull its supply voltage below the operating range.
The firmware for the setup, as well as the scripts used for the final attacks, are available in the artifacts (\autoref{sec:open-science}).

\subsection{Integer Overflow}
\label{sec:integer_overflow}
Without causing any glitches, we found that Macronix, Winbond, and GigaDevice chips are susceptible to integer overflows.
The counter value appears to be stored internally as a 32-bit unsigned integer.
If a counter with a value of 4294967295 (the maximum value of a 32-bit unsigned integer) is incremented once more, it returns to 0.
This enables attackers to issue authentic \rpmcinc commands to set it to lower values, contradicting the implicit assumption that the counter should be ``monotonic''.

\paragraph{Duration}
In our testing, it took about 4 weeks to overflow the Winbond counter and about 6 weeks for the Macronix or GigaDevice chip.
Theoretically, the average time for an increment request is 80\textmu s in Winbond's \cite{winbond_w25r128jv} and GigaDevice's \cite{gigadevice_gd25lr256e} case or 45 \textmu s in Macronix's case \cite{macronix_mx77l12850f}.
This means a full overflow cycle could be as fast as 4 or 2.5 days, respectively.
This theoretical estimate, however, does not account for the time required to transmit the necessary messages or the possibility of transactions failing, which occurred sporadically during our testing.
Nonetheless, significant speedups could be achieved by precomputing signatures on more powerful hardware.

\paragraph{Mitigations}
Despite allocating 4 bytes, i.e., 32 bits, for the counter value, the RPMC specification fails to specify the expected behavior of increments to the maximum integer \cite{jesd260}.
This contrasts with the specification of RPMB within eMMC, where it is directly stated that a counter should reject all increments once it reaches its maximal value \cite{jesd84}.
JEDEC could resolve this by either deciding on a single behavior and documenting it, or by using one of the reserved bits in the \rpmcroot command to make this behavior configurable for the end user.
We contacted JEDEC with this proposal in April 2025, but they have not updated the specification since then.

\paragraph{Impact}
We believe this fault is not caused by a single manufacturer's implementation but rather by ambiguity in the RPMC specification.
It forces manufacturers to make a choice, and leaves end users in the dark about exactly how their chip behaves and whether it meets their security requirements.
For the use case of NV state rollback protection, the ability to overflow a counter and arrive at a lower value than before is only of marginal use:
It could allow one-time reuse of stale NV state, but would require another overflow to reuse it.
The impact would be much higher, however, if RPMC is used as a ratcheting mechanism to protect a device's firmware against rollback attacks, using a pair of software and hardware version numbers.
Once the counter overflows, all previous firmware versions become valid again. This breaks ratcheting's security goal to prevent firmware rollbacks.

\subsection{Counter Corruption}
We found the GigaDevice flash to be susceptible to voltage glitching during the \rpmcget process.
With the right timing, we were able to get the chip to return faulty counter values.
The returned values are indistinguishable from valid responses, as the tag matches the one requested by the controller and a valid signature signs the whole message (covering the faulty counter value). 
During glitching, we additionally observed the chip sometimes returning a different tag or a faulty signature.
This is not a direct vulnerability, as any well-implemented controller should easily catch and reject these.

\paragraph{Parameters}
With a well-configured setup, we were able to trigger the glitch with a success chance of around 10\%.
The best working MOSFET open time was closing it with no delay of 0 or 140 ns while glitching 10500 (roughly 200\textmu s) to 14300 (roughly 240 \textmu s) decrements after driving the chip select of the \rpmcget request high again.
The later we glitched, the higher the wrong counter value would be, but the glitched value would still stay below the real counter's value.

\paragraph{Mitigations}
As we have observed this behavior only on GigaDevice flash chips, we believe the best defense against this attack is a hardware revision of those chips that better protects the loading process of the counter against fault injection.
Another approach would be to insert standard fault-injection countermeasures to detect voltages outside normal operating conditions and subsequently reset or stall the chip's operation to prevent abnormal behavior.

\paragraph{Impact}
The real-world impact of this vulnerability is low for the use cases we observed.
Since the returned counter values still exhibit some randomness, one would need to build a payload that specifically matches one of them.
Since this would probably require multiple glitch attempts, the attacker needs to keep the application waiting by returning a busy status.
But even then, for every \rpmcinc validated with \rpmcget, this process would need to be repeated until you glitch a value exactly one higher.
This was impractical for the real-world attack targets we found, but could still impact other use cases we have not seen so far. 

\subsection{HMAC Key Leak}
\label{sec:hmac_key_leak}

We found the Winbond flash chips to be susceptible to voltage glitching at two critical points:
After issuing an \rpmcinc command, we applied a crowbar glitch during what we think is the chip's signature verification operation.
When we issued a subsequent \rpmcstatus, it would occasionally return a valid signature for our initially sent \rpmcinc, even when the initial request lacked a valid signature.
More critically, when we encountered a similar glitch after an \rpmcupdate command, the chip returned the calculated HMAC key in a subsequent \rpmcstatus, regardless of whether our initial \rpmcupdate was signed correctly.

Winbond seems to reuse the register read by \rpmcstatus to store the chip's HMAC engine output.
A well-timed voltage glitch then appears to stop the process that would normally clear this register again, allowing its output to leak.
We have also tested both chips from Winbond's 3.3V and 1.8V lines and even performed this attack against the flash chip in a commercially available laptop.
All attacks were successful, leading us to conclude that this is a design flaw in Winbond's RPMC engine.

\paragraph{Parameters}
Using the best parameters we found, we can leak the HMAC key within roughly 1 second, enabling even on-demand glitching with a purpose-built interposer.
We found a MOSFET open delay of 0 to 5 (roughly 140~ns--160~ns) to work best, and a glitch with a delay of around 7300--7350, roughly 340 \textmu s after the initial chip select low of the request.

\paragraph{Mitigations}
\label{sec:winbond-mitigations}
Winbond's datasheet even reproduces JEDEC's own requirement almost verbatim: ``The increment counter implementation should make sure that the counter increment operation is performed in a power glitch aware manner.'' \cite{jesd260, winbond_w25r128jv}.
We believe the only way to remedy the fault described in this section is through a hardware revision to Winbond's RPMC engine.
This hardware revision should address register reuse and ensure that no secret data is ever written to a register that can be read.

\paragraph{Impact}
This vulnerability severely impacts any application that relies on Winbond flash chips and is almost equivalent to leaking the \emph{Root Key} from such a chip.
With the HMAC key, an attacker can use an interposer to sign their own responses to \rpmcget and fill them with arbitrary counter values that benefit them.
This fully nullifies any security within RPMC.
Since the glitch is also very fast, glitching can be performed on demand once the \textit{Key Data} for the current session is known. 
There is no practical way to detect this, as the attacker could return a busy status from the chip until the glitch succeeds.
Glitching that causes a timeout is unlikely, and in all the real-world use cases we saw, timeouts were not an issue.
But even if it occurs, the attacker could extend the time they have by performing the attack during the period from the \rpmcupdate to the first \rpmcget, as this is the first request that actually requires a signed status.

\section{Interposer Attacks}
\label{sec:interposer-attacks}

\begin{table*}[h]
    \centering
    \caption{Evaluated Target Systems. Integration patterns of RPMC are unclear to us.}
    \small
    \begin{tabular}{lllllll}
        \toprule
        \textbf{Device} & \textbf{Year} & \textbf{Platform} & \textbf{CPU} & \textbf{Flash} & \textbf{uses RPMC} & \textbf{Microsoft branding} \\
        \midrule
        Intel NUC Desktop Board NUC6i3SYK & 2015   & Intel & Core i3-6100U         & Winbond W25Q64FVSIG         & No  & --              \\
        ASUS TUF Gaming B450-Plus         & 2019   & AMD   & Ryzen 5 3600 (Zen~2)  & Winbond W25Q256JWEQ         & No  & --              \\
        Lenovo IdeaPad 5 Pro 16ACH16      & 2021   & AMD   & Ryzen 5600H (Zen~3)   & Winbond 74M12JWSIQ          & Yes$^\ast$ & --       \\
        Lenovo ThinkPad T15 Gen 2                & 2021   & Intel & Core i7-1165G7        & Winbond W25R256JVEN         & Yes & Secured-core PC \\
        MSI PRO B760M-P DDR4              & 2024   & Intel & Core i5-14400         & Winbond W25Q256JWEQ         & No  & --              \\
        MSI PRO B850-P WIFI               & 2024   & AMD   & Ryzen 5 9600X (Zen~5) & Macronix MX25U25671G        & No  & --              \\
        Framework Laptop~13               & 2024   & Intel & Core Ultra~5 125H     & Winbond 25R256JW            & Yes & --              \\
        HP OmniBook~3 15-fn0523ng         & 2025   & AMD   & Ryzen AI~5 330        & GigaDevice 25LR256FYIGe     & Yes & Copilot+        \\
        Acer Aspire 14 AI OLED            & 2025   & Intel & Core Ultra~5 226V     & Macronix MX25U25672G        & No  & Copilot+        \\
        \bottomrule
    \end{tabular}
    \\[2pt]
    \raggedright\footnotesize $^\ast$Winbond markets the W74M series as \emph{Authentication Flash} \cite{winbond_w74M12jw}; it uses the same commands as RPMC flash, and we observed these commands on the SPI bus during boot.
    \label{tab:evaluated_systems}
\end{table*}

In \autoref{sec:flash_vulnerabilities}, we evaluated attacks against isolated RPMC-enabled SPI flash chips.
Especially the \emph{HMAC Key Leak} (\autoref{sec:hmac_key_leak}) should allow an attacker to circumvent RPMC-backed security mechanisms such as NV state rollback prevention (see \autoref{sec:bg_ftpms}).
However, implementing such an attack requires non-trivial technical effort, including overcoming electronic limitations, operating at high SPI clock speeds, and performing manufacturer-specific reverse engineering.
In this section, we present our solution to these challenges: our custom-made \textit{interposer PCB} (\autoref{sec:interposer_pcb}).
\\
Furthermore, when investigating AMD’s and Intel’s NV state rollback protection in detail, we discovered security issues with Intel's (\autoref{sec:hmac_key_reuse}) and AMD's (\autoref{sec:unverified_increment}) use of RPMC-enabled SPI flash chips.
While Intel’s weakness will ease the counter forgery enabled by Winbond’s \emph{HMAC Key Leak} (\autoref{sec:hmac_key_leak}), AMD’s flaw will allow us to attack systems regardless of whether the SPI flash contains vulnerabilities or not.

\subsection{Attacker Capabilities}
\label{sec:interposer-attack-capabilities}

To attack a system using a vulnerable RPMC chip, an attacker can leverage their \textit{physical access} and manipulate, e.g., the chip's response to a \rpmcstatus command.
More concretely, an active attacker needs the following capabilities:

\begin{description}[labelindent=1em, leftmargin=2em]
  \item[Eavesdrop] on the RPMC communication between the controller and SPI flash chip to keep track of the state of the controller and chip, and to trigger manipulations at the appropriate times.
  \item[Manipulate] the RPMC traffic between the controller and SPI flash chip, specifically:
    \begin{itemize}
        \item Filter RPMC commands sent from the controller towards the SPI flash chip.
        \item Respond to these commands with attacker-chosen responses signed with a leaked HMAC key.
    \end{itemize}
  \item[Glitch] the SPI flash chip to leak the necessary HMAC keys (optional due to \ref{sec:hmac_key_reuse}).
\end{description}

These capabilities align with the assumed threat model specified in the RPMC standard \cite{jesd260}.

\subsection{Attack Setup}
\label{sec:interposer_pcb}

\begin{figure}
    \centering
    \includegraphics[width=0.95\linewidth]{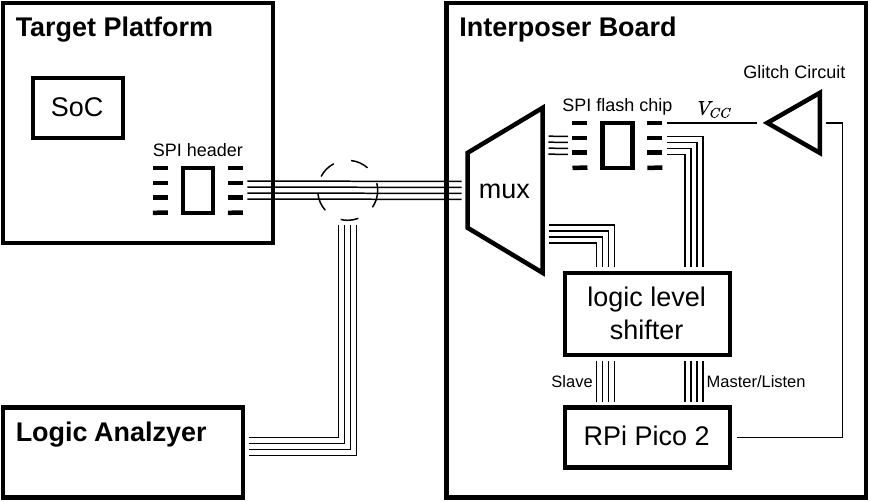}
    \caption{Attack Setup with the Interposer Board}
    \Description[A block diagram showing the interposer board wired between a target platform's SoC and its SPI flash chip]{The target platform's SoC connects over its SPI header to a mux on the interposer board, which switches the bus between the original SPI flash chip and an RPi Pico 2 acting as a listening/spoofing SPI device through a logic level shifter; a glitch circuit can pull the flash chip's V\_CC, and a logic analyzer taps the SPI header for monitoring.}
    \label{fig:interposer-board}
\end{figure}

\begin{figure}
    \centering
    \includegraphics[width=0.95\linewidth]{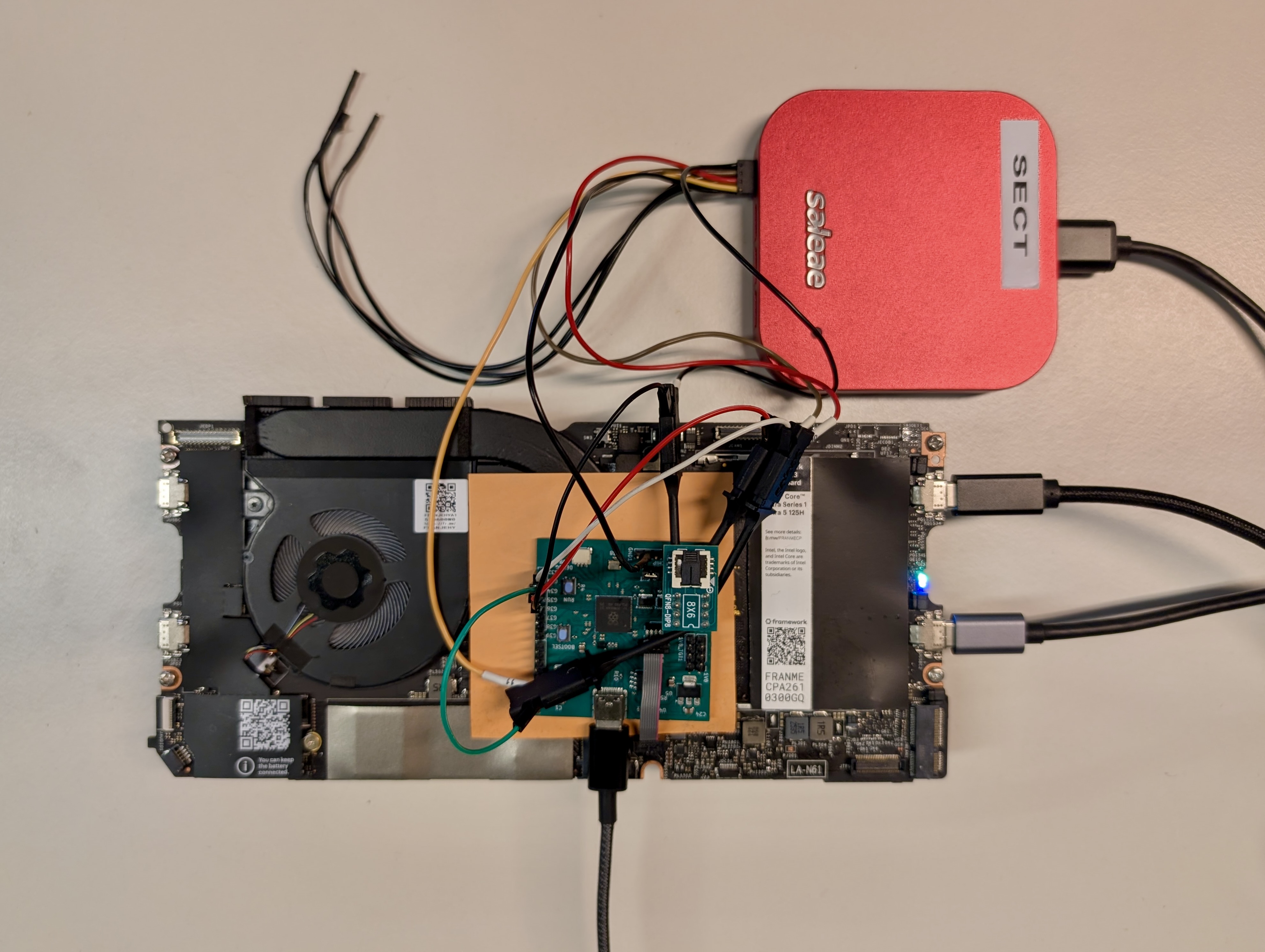}
    \caption{Photo of the Attack Setup against the Framework Laptop 13's mainboard}
    \Description[A picture showing the interposer and a logic analyzer connected to a framework laptop mainboard]{}
    \label{fig:interposet-board-photo}
\end{figure}

Manipulating selected SPI transactions on a high-speed SPI connection is technically challenging.
To facilitate this manipulation, we desolder the target's SPI flash chip and place our interposer PCB between the controller and the SPI flash chip.
This setup must minimize interference with the SPI bus, as QSPI traffic can reach clock speeds up to 100 MHz.
It must also listen in on the SPI traffic and trigger manipulations of this traffic with a latency of only dozens of nanoseconds.

\paragraph{Hardware}
At the core of this PCB lies a QSPI-capable MUX IC \cite{texasinstruments_TS3A27518E}, which -- for most SPI traffic -- enables a nearly-direct connection between the target platform and the de-soldered SPI flash chip.
When an SPI command should be manipulated, the MUX is configured to connect the target's SPI bus not to the SPI flash chip, but to a Raspberry Pi (RPi) Pico 2W's SPI ``slave'' interface \cite{rpi_pico_2w}.
The RPi Pico 2W will respond to the controller's SPI commands instead of the SPI flash chip.
Additionally, the RPi Pico 2's other SPI interface is permanently connected to the SPI flash chip's SPI bus to monitor SPI traffic and take over the controller role during manipulation phases.

Due to varying target SPI bus voltage requirements, a \textit{Logic Level Shifter} (LLS) IC \cite{74LVC16245A} translates the $3.3 V$ signals of the RPi Pico 2 to the corresponding target voltage (most often $1.8 V$).
We also connect a logic analyzer to observe and debug the interposer logic. 
A MOSFET on the SPI flash chip's voltage rail enables in-circuit voltage-fault injection attacks on the SPI flash chip.

To insert the interpose PCB between a target and its SPI flash chip, we desolder the SPI flash chip and attach a $1.27$ mm pin header instead.
The interposer can then be connected directly to this header, and the SPI flash chip mounted in a socket fitting its package.
See \autoref{fig:interposet-board-photo} for an exemplary hardware setup.

\paragraph{Firmware}
The RPi Pico 2W was chosen for its low price -- around $5\$$ -- and its RP2350 chipset, which includes \textit{Programmable Input Output (PIO)} blocks \cite{rp2350_datasheet}.
These allow us to implement time-critical attack logic as low-latency state machines, without the need for an (expensive) FPGA.

We implemented a PIO state machine to listen to the target's SPI traffic and trigger on the first bytes of an SPI command, which identifies its type.
This is then used to trigger other parts of the attack logic, such as functionality to record and process the current SPI command with the RPi Pico 2W's firmware.
Once the firmware determines that future RPMC commands should be manipulated, it can also configure this state machine to instead switch the MUX over to the RPi Pico 2W's SPI ``slave'' interface after such a trigger.
The PIO state machine enables such a manipulation decision within only a few dozen nanoseconds.
Additionally, another state machine on the same PIO hardware can be used to output a firmware-determined bit sequence on the SPI bus' MISO line, enabling the firmware to respond to the controller's RPMC commands.

\paragraph{Board Revision}
\label{sec:interposer_v2}
The board described above (v1) suffered from signal-integrity and timing issues on the interposed QSPI bus at the Framework Laptop 13's operating SPI clock speed, preventing reliable command manipulation against that target.
Since the Winbond/Intel attack path (\autoref{sec:nv_rollback_leaked_key}) only needs to \emph{replay} an HMAC key already leaked once via voltage glitching (\autoref{sec:hmac_key_leak}) -- rather than glitch on demand -- we built a second, replay-only board revision (v2) to address this.
Compared to v1, v2 (a) configures the RPi Pico 2's GPIO bank for native $1.8 V$ I/O from the start, removing the Logic Level Shifter entirely; (b) omits the glitching MOSFET, since no fresh key extraction is needed; and (c) uses a faster MUX IC \cite{texasinstruments_TMUX1574DYY}.
Removing the LLS and glitching circuitry simplified the signal path, which, combined with the faster MUX, resolved the timing and signal-integrity issues that had prevented reliable command manipulation against the Framework Laptop 13.
We used v2 only for the Intel/Framework result (\autoref{sec:nv_rollback_leaked_key}); the AMD/OmniBook result (\autoref{sec:nv_rollback_increment_suppression}) stands as originally demonstrated on v1.

\subsection{Naive Rollback Attempts}
\label{sec:nv_state_rollback_protection}

To understand the vulnerabilities and attacks that follow, we will now revisit a critical detail when working with security applications that require non-volatile storage, such as fTPMs.
An fTPM usually stores its NV state alongside the firmware on the SPI flash chip and protects its integrity and confidentiality using well-known cryptographic methods.
However, to reliably implement anti-hammering mechanisms (\autoref{sec:bg_anti_hammering}), it must also prevent rollback attacks against the NV state.

Of the nine systems we examined during our research (see \autoref{tab:evaluated_systems}), four support RPMC (including Winbond's functionally identical \emph{Authentication Flash}) and five do not.
On AMD-based systems, the NV state (including the fTPM state) is contained in a file of the PSP file system \cite{faultpm_2023}. On Intel-based systems, the NV state is part of the ME File System inside the CSME Region of the UEFI volume \cite{sklyarov_mfs_explained}.
\\
We conducted our test on systems with the fTPM enabled in the UEFI and a TPM-protected secret under dictionary attack protection, i.e., either a Windows Hello PIN and BitLocker on Windows 11 or a TPM object created with \emph{tpm2-tools} on Linux.
The test comprised these steps:
\begin{enumerate}
    \item Attach an SPI programmer, dump the SPI flash, and detach.
    \item Boot up the system, change the fTPM state, e.g., by increasing the TPM's lockout counter through a failed authentication attempt, and shut down cleanly.
    \item Attach the  SPI programmer, write the dumped SPI flash image, and detach.
    \item Boot up the system and check whether the fTPM state was rolled back.
\end{enumerate}

We distinguished successful from unsuccessful rollbacks on either operating system.
On Windows, a BitLocker recovery screen on boot indicates a reset fTPM, and the \code{Get-Tpm} PowerShell command reports the exact \code{LockoutCount}.
On Linux, \code{tpm2\_getcap properties-variable} reports the lockout counter, and loading a previously created TPM object fails if the fTPM was reset.

On seven of the nine systems, the fTPM detected the rollback.
On Windows, instead of booting directly, we were asked to enter the BitLocker recovery key; after Windows booted, an error message said the Windows Hello PIN was unavailable, forcing us to set up Windows Hello again.
On Linux, loading our TPM object failed its integrity check.
Both symptoms indicate that the fTPM no longer had access to its previous objects.
These seven platforms (with or without RPMC support) must have detected the stale NV state through their mismatches between a counter in the (encrypted and integrity-protected) NV data and an expected value saved elsewhere.
They therefore rejected using it for the boot and, instead, reset its contents to allow the system to boot with a fresh fTPM.

Both AMD desktops (Zen~2 and Zen~5), whose flash lacks RPMC, did not detect the rollback.
On these systems, reprogramming an old NV state onto the SPI flash with an SPI programmer alone reliably returned the fTPM's lockout counter to its value at the time of the flash dump.

\textbf{It is important to note that the Intel-based ``pre-RPMC'' systems have defended this naive attack.}
While the CSME whitepaper mentions a \emph{Protected Real-Time Clock} (PRTC) that could enable the rollback protection~\cite{csme_whitepaper}, independent research indicates that the counter values of the ME File System's metadata must match a so-called Anti-Replay table (AR)~\cite{sklyarov_mfs_explained}.
Regarding the rollback protection of this table, Sklyarov states: ``RPMC is an optional feature of the SPI Flash chip helpful for Anti-Replay Protection implementation. If this feature is not available, ME implements its own timer-based counter.''~\cite{sklyarov_mfs_explained}
Our naive rollback tests on the Intel systems without RPMC flash -- the NUC from 2015, the MSI desktop, and the Acer Aspire from 2024 and 2025 -- support this claim.
For AMD, our results indicate the opposite:
The Lenovo IdeaPad, which we initially classified as pre-RPMC, relies on Winbond's \emph{Authentication Flash}, and both AMD systems without such flash accepted naive rollbacks.
While these observations may not be representative of all AMD systems, we conclude that AMD SoCs likely do not feature standalone rollback protection for their NV state.

\subsection{Controller Vulnerabilities}
\label{sec:controller_vulnerabilities}

In order to better understand how AMD and Intel protect their NV state on the four systems \emph{with} RPMC support (\autoref{tab:evaluated_systems}), we collected traces\footnote{The target systems  (\autoref{tab:evaluated_systems}) for these traces were the OmniBook 3 with PSP firmware version: 0.3E.7.79 and the Intel-based Framework Laptop 13 with CSME firmware version: 18.0.15.2518} of the SPI bus during early boot and during runtime, namely a failed Windows Hello PIN entry, whose anti-hammering protection is backed by the fTPM (see \autoref{sec:windows_hello}).

Our key observations from the traces were these (highlighted in bold are the ones critical for the upcoming controller vulnerabilities): during early boot, AMD and Intel both initialize the first four RPMC counters via \rpmcupdate, and \textbf{Intel calls \rpmcupdate with the same \textit{Key Data} value of 0x1 across all boots}; AMD additionally calls \rpmcupdate on counter 0 a second time; and while AMD only ever uses counter 0 afterwards, Intel uses both counter 0 and counter 1.
After every \rpmcinc, \textbf{AMD does not verify the counter via an \rpmcget}, which in turn Intel does.

\subsubsection{HMAC Key Reuse}
\label{sec:hmac_key_reuse}

The \textit{Key Data} field should be used to generate a unique HMAC key each time the \rpmcupdate command is issued \cite{jesd260}.
The finding that Intel-based systems use the constant \textit{Key Data} value 0x1 across reboots is a security weakness that is not exploitable in isolation, as it only slightly reduces the replay-attack resilience of the RPMC specification.
\footnote{
An attacker who has recorded $2^{48}$ \rpmcget responses is expected to succeed in maliciously answering a later \rpmcget command with a recorded response after $2^{48}$ attempts (assuming a constant \textit{Key Data} field, but randomly chosen 96-bit \textit{Tag} values).
For a randomly chosen \textit{Key Data} field, the most efficient attack requires $2^{64}$ recordings and is expected to succeed after $2^{64}$ malicious answer attempts.
}

However, this finding significantly reduces the attack efforts necessary to turn Winbond’s \emph{HMAC Key Leak} (\autoref{sec:hmac_key_leak}) into an NV state rollback attack:
Instead of performing the voltage fault injection attack each time an \rpmcupdate command is issued, the constant \textit{Key Data} value allows us to perform the voltage fault injection only once.
The leaked HMAC key for the constant \textit{Key Data} value of 0x1 can then be used to forge all \rpmcget replies across all boots. 
The obvious mitigation for this issue is to choose \textit{Key Data} securely randomly for each \rpmcupdate command.

\subsubsection{Unverified Increment}
\label{sec:unverified_increment}

Our RPMC traces surfaced another, more critical security issue on AMD-based systems.
Recall that the only response from an RPMC-enabled SPI flash chip that is cryptographically signed is the \rpmcget response \cite{jesd260}.
On AMD, \rpmcinc is the only RPMC command triggered by the fTPM-backed Hello PIN entry.
This means that the AMD SPI controller does not receive any confirmation about the increment success.
An attacker can simply filter out this \rpmcinc request so that the increment operation is never performed on the SPI flash chip, and send a malicious response that appears to confirm a successful increment without the ability to sign responses.
This allows an attacker to roll back the NV state on RPMC-backed systems, as outlined in \autoref{sec:nv_rollback_increment_suppression}.

We view this attack vector as particularly impactful, as it works with every RPMC-enabled SPI flash chip, regardless of whether the chip is vulnerable to our voltage-fault attack.

\label{sec:amd_mitigations}
On a positive note, AMD should be able to fix this vulnerability through firmware updates:
To protect against this attack, RPMC controllers must verify any allegedly successful increment operations by re-reading the counter value, as Intel's RPMC usage already does by calling \rpmcget after each \rpmcinc.

\subsection{RPMC-based NV State Rollback Attack}
\label{sec:nv-state-rollback-attack}

Utilizing the vulnerabilities discussed in Sections \ref{sec:hmac_key_leak}, \ref{sec:hmac_key_reuse}, and \ref{sec:unverified_increment}, and our custom interposer (\ref{sec:interposer_pcb}), we implemented two attack paths against AMD's and Intel's RPMC-backed NV state rollback protection mechanisms.

\subsubsection{Suppressing Increments}
\label{sec:nv_rollback_increment_suppression}

\begin{figure*}[h]
    \centering
    \includegraphics[width=1.0\linewidth]{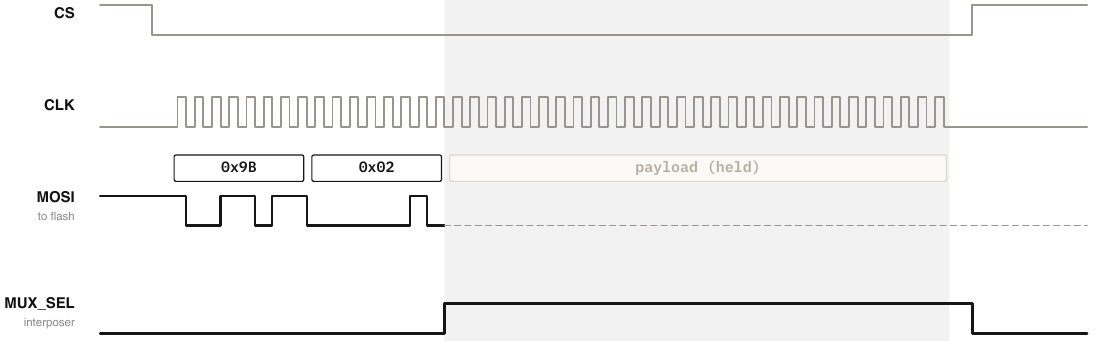}
    \caption{Selective suppression of an \rpmcinc (0x9B 0x02) command}
    \Description[A logic-analyzer timing diagram of the SPI CS, CLK, and MOSI lines showing the interposer holding back an Increment Monotonic Counter command's payload]{Timing traces for chip-select, clock, and MOSI show the 0x9B 0x02 opcode bytes of an Increment Monotonic Counter command being sent normally, after which a shaded region marks the interposer's MUX\_SEL line switching over and holding the remaining payload bytes so they never reach the flash chip.}
    \label{fig:spi_suppress_increment}
\end{figure*}

We attacked the AMD-based \textit{HP OmniBook 3 15-fn0523ng} system (see \autoref{tab:evaluated_systems}) by attaching our interposer PCB between the system's motherboard and the desoldered SPI flash chip. 
This allowed us to program the RPi Pico 2W to intercept any \rpmcinc commands and prevent them from reaching the SPI flash chip.
We use a PIO state machine to detect any \rpmcinc commands and disconnect the controller from the SPI flash chip during said command (see \autoref{fig:spi_suppress_increment}).
Since only the command byte is transmitted, the command appears as an invalid command to the SPI flash chip, so no actual increment is performed.
After detecting a \rpmcinc command, the RPi Pico 2W will respond to any \rpmcstatus commands with a success response, causing the AMD system to assume the increment was successful.
Finally, we stop intercepting SPI traffic so the flash chip can continue to respond to the controller's requests.

\subsubsection{Using Leaked Key}
\label{sec:nv_rollback_leaked_key}

The Intel system (\textit{Framework Laptop 13} in \autoref{tab:evaluated_systems}) with a vulnerable \textit{Winbond 25R256JW} chip was also modified by adding our interposer PCB (v2, \autoref{sec:interposer_v2}) between the motherboard and the SPI flash chip.
For the attack, the interposer PCB must track all RPMC commands and issue manipulated responses to \rpmcget and \rpmcinc.
These manipulated responses aim to reflect the state of an RPMC-enabled SPI flash chip, but with attacker-chosen counter values.

Due to Intel's constant \textit{Key Data} value (\ref{sec:hmac_key_reuse}), we did not need to re-extract the HMAC session keys after every \rpmcupdate, but could hard-code them for each counter after a one-time voltage fault injection attack (\ref{sec:hmac_key_leak}).
The interposer PCB's firmware then uses a PIO state-machine to detect \rpmcget commands.
For each, it waits for the following \rpmcstatus command -- while letting through any other, non-RPMC SPI commands -- and uses that time to fabricate a response containing a chosen counter value, and sign it with the leaked HMAC key.
As soon as the interposer detects a \rpmcstatus command, it disconnects the flash chip from the target and returns its own forged response.
Afterward, the interposer reconnects the target and SPI flash chip to ensure reliable (non-RPMC) SPI communication.

The interposer's firmware also listens for \rpmcinc commands to keep track of the spoofed counter values.
Additionally, it also interposes all immediately following \rpmcstatus commands as described above, because otherwise the flash chip would indicate the counter value mismatch in the status, since the current expected value is included in the \rpmcincshort command (\autoref{sec:rpmc_cmd_increment}).

\subsubsection{Results}
\label{sec:interposer_results}

Both attack paths reliably trick their respective target into accepting an expired NV state during boot, thereby circumventing the RPMC-backed rollback prevention: the AMD suppression path (\autoref{sec:nv_rollback_increment_suppression}) on the OmniBook, and the Winbond leaked-key path (\autoref{sec:nv_rollback_leaked_key}) on the Framework Laptop 13 using the v2 interposer board (\autoref{sec:interposer_v2}).
We build on the AMD path for the end-to-end Windows Hello attack (\autoref{sec:hello_bruteforce}) and on the Winbond path for the end-to-end BitLocker attack (\autoref{sec:bitlocker-bruteforce}); note that this pairing is arbitrary, since the underlying NV-state rollback primitive is independent of the target application -- either attack path could equally be paired with either application.
Furthermore, we expect the suppression-based attack path to work the same way when Microsoft's Pluton TPM is enabled, as we observed the same RPMC usage with unverified SPI bus traffic increments (\autoref{sec:controller_vulnerabilities}).

\section{Windows Hello and BitLocker PIN Brute-Forcing}
\label{sec:hello_bruteforce}

In \autoref{sec:interposer-attacks}, we presented our custom interposer boards that we can
use to modify RPMC transactions using different vulnerabilities to rollback the
NV state. We will now describe how we used these boards in two end-to-end attacks.
First, we brute-forced the Windows 11 Hello PIN on locked and encrypted
Copilot+ hardware, specifically the HP Omnibook (see
\autoref{tab:evaluated_systems}), using the AMD \emph{Unverified Increment} rollback path.
The same rollback primitive is not specific to Hello: \autoref{sec:bitlocker-bruteforce} describes our second end-to-end attack, live BitLocker PIN brute-forcing on the Framework Laptop 13 using the Winbond \emph{HMAC Key Leak} rollback path, which defeats BitLocker's TPM+PIN pre-boot protector even more effectively.

\paragraph{Attacks Overview}

Assuming a physical attacker with no knowledge of a user's PIN, this section outlines three different brute-force attacks against Windows Hello.
The first, theoretical, attack describes a naive brute-force attack on Hello PINs, limited by lockout delays imposed by both Windows and the TPM. %
It serves as the baseline for estimating the impact of the two following attacks.

The second attack assumes that the attacker is able to conduct an NV State rollback between reboots (as outlined in \autoref{sec:nv_rollback_increment_suppression} and \autoref{sec:nv_rollback_leaked_key}).
As experimentally verified by us, it demonstrates brute-forcing without any lockout durations enforced by Windows \emph{or} the TPM.
However, it is still limited to 5 tries per boot.

Lastly, we evaluate a theoretical third attack where an attacker obtains access to the sealed TPM object that Windows sends to the TPM for decryption.
While this attack scenario requires at least temporary file system access, we argue that it can serve as a privilege-escalation technique.
Fundamentally, it would allow an attacker to increase their tries per boot from 5 (Windows-enforced) to 32 (TPM-enforced).

\paragraph{Lockout Configurations}
\label{sec:bruteforce_lockout_conf}

We noticed that lockout configurations differ from Microsoft's documentation and among our tested devices (\autoref{tab:evaluated_systems}).
The \emph{Windows Hello for Business} FAQs claim that Hello provides four PIN entries per reboot (with a ``A1B2C3'' captcha after the third try) \cite{whfb_faq}, while our target systems (with regular Hello set up) allowed one more try before the captcha; so a total of five tries.
It further describes increasing lockout durations of ``1 minute after the first reboot, 2 minutes after the fourth reboot, and 10 minutes after the fifth reboot'' \cite{whfb_faq}.
While we confirmed that the HP OmniBook employs this style of lockout duration after reboots, we did not see it on the Acer Aspire.

Running \texttt{Get-TPM} in PowerShell reveals the TPM's actual lockout counters.
To our understanding, the lockout counters need to reach 32 so an actual TPM-enforced lockout would apply.
On the OmniBook, the firmware's factory-provisioned \emph{LockoutHealTime} is 10 minutes, as documented by Microsoft as the default; however, once Windows 11's Hello PIN setup completes, this changes to 2 hours, matching what we observed on the Acer Aspire and the Framework Laptop but contradicting Microsoft documentation \cite{microsoft_tpm_fundamentals}.

\subsection{Naive Brute-Forcing}

We first assume an attacker who does not have access to the TPM object used during the login process because the hard drive is BitLocker-encrypted (see \autoref{sec:bg_disk_encryption}). 
Thus, instead of verifying their attempts directly with the TPM, they need to interact with the Windows login screen and check the PINs there.

\paragraph{Results}

\begin{table}[h]
  \centering
  \caption{Worst-Case Numeric PIN Retrieval Time, Naive Brute-Force vs.\ Our Measured NV Rollback Attack: for Windows Hello, 69 days collapse to 40 hours; for BitLocker's TPM+PIN Protector, 19 years collapse to 20 days. NV Rollback bypasses \emph{recoveryTime} entirely and is therefore unaffected by it.}
  \small
  \label{tab:bruteforce-speeds}
  \renewcommand{\arraystretch}{0.9}
  \begin{tabular}{lllll}
    \toprule
    & \multicolumn{2}{c}{Naive (with lockout)} & \multicolumn{2}{c}{\textbf{with NV Rollback}} \\
    \cmidrule(lr){2-3} \cmidrule(lr){4-5}
    Digits & 2 h & 10 min & \textbf{Hello} & \textbf{BitLocker} \\
    \midrule
    4 & 833 days    & 69 days   & \textbf{40 hours} & 5 hours$^\dagger$ \\
    6 & 228 years   & 19 years  & 168 days & \textbf{20 days} \\
    8 & 22831 years & 1903 years & 46 years & 6 years \\
    \bottomrule
  \end{tabular}
  \\[2pt]
  \raggedright\footnotesize $^\dagger$Only reachable if an administrator explicitly lowers the default PIN length floor.
\end{table}

We calculated the worst-case brute-force times for the naive brute-forcing in \autoref{tab:bruteforce-speeds}, once for the 10-minute \emph{recoveryTime} that Microsoft documents as the default, and once for the 2-hour \emph{recoveryTime} we observed on the Acer Aspire and the Framework Laptop (\autoref{sec:bruteforce_lockout_conf}).
Notably, with the 10-minute setting, a four-digit PIN can be recovered after 69 days; with the 2-hour setting, the same worst case grows to 833 days, a 12$\times$ increase proportional to the \emph{recoveryTime} ratio.
We think that this security level (69 days worst-case brute-force for a TPM-backed 4-digit PIN) can already be inadequate for certain users.

\paragraph{Mitigations}

While enterprises may use Microsoft's WHFB policy settings to increase the minimum PIN length or require letters \cite{whfb_policy_settings}, they do not allow changing the 10-minute recovery delay after 32 failed attempts.
We believe Windows should prevent naive brute-forcing as just described by making the 32 try limit a \emph{permanent} limit and use this policy by default:
It is highly unlikely that users who already entered their PIN incorrectly 32 times will recall their correct PIN on the 33rd attempt (enabled after the TPM-configured \emph{recoveryTime}  (\ref{sec:bg_anti_hammering})).
After all, Microsoft provides a flow to reset the PIN using the user's Microsoft account and an internet connection.
Enforcing a permanent lockout would match the behavior of, e.g., iPhones \cite{apple_platform_security}.

\subsection{Brute-Forcing with NV Rollbacks}
\label{sec:brute-forcing-with-nv-rollbacks}

In the second attack, we assume an attacker again without access to the TPM object, who is therefore limited to interacting with the Windows login screen, as in the previous scenario.
However, this time they can also rollback the NV state during rebooting using our interposer board (\ref{sec:nv-state-rollback-attack}).

\paragraph{NV State Rollback}

Assume the NV state rollback primitive from \autoref{sec:nv-state-rollback-attack}.
We mounted the same attack on our OmniBook, but before booting, started an interposing logic appropriate for the attack.
For AMD, this means the interposer suppresses all \rpmcinc commands (\ref{sec:nv_rollback_increment_suppression}).
Additionally, the interposer has access to an initial dump of the PSP NV state.
During reboot, it issues an SPI write to the flash chip, overwriting any current NV state with the dumped state.

\paragraph{KVM Automation}

\begin{figure}
    \centering
    \includegraphics[width=0.95\linewidth]{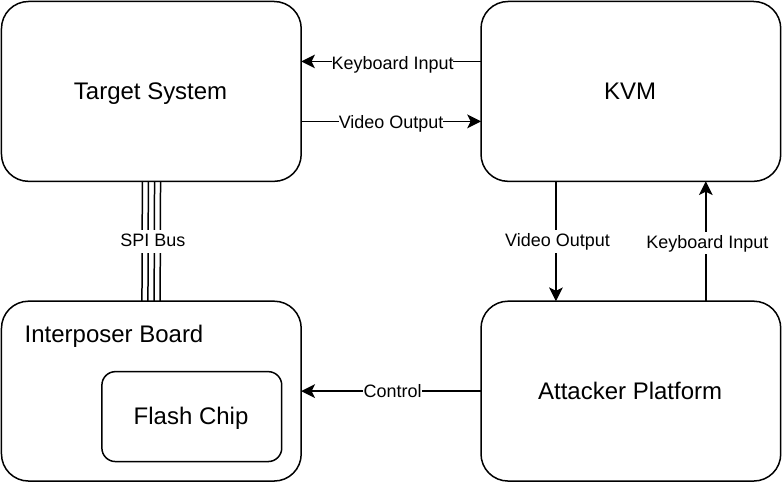}
    \caption{Brute-force Setup with the Interposer Board and a KVM}
    \Description[A block diagram of the automated brute-force loop connecting the target system, a KVM, an attacker platform, and the interposer board]{The target system exchanges keyboard input and video output with a KVM, which forwards video to and receives keyboard commands from an attacker platform; the same attacker platform controls the interposer board, which sits on the SPI bus between the target system and its flash chip to roll back the NV state between reboots.}
    \label{fig:bruteforce-setup}
\end{figure}

To brute-force the Hello PIN via the Windows login screen, we have to connect, in addition to our interposer board, a KVM to the target system (see \autoref{fig:bruteforce-setup}).
It allows us to read the video output and trigger keyboard input.
The KVM is further controlled by an attacker platform that uses OCR to determine the state of the login screen.
It then sends keystrokes through the KVM to the target to input a PIN or trigger a reboot once Windows mandates it (i.e., after 5 tries, see \autoref{sec:windows_hello}).
During the reboot, our board reprograms the NV state to its previous state.

In detail, the automation flow is this:
\begin{enumerate}
    \item Boot and wait for the lock screen
    \item Enter PIN (4x)
    \item Enter Captcha (``A1B2C3'')
    \item Enter one more PIN
    \item Restart (using keystrokes)
    \item Reprogram flash chip
    \item Repeat
\end{enumerate}

Once the correct PIN is found, the Windows desktop appears, and our KVM logic halts. 

\paragraph{Results}

After reboot, we observed that the system accepted the next 5 PIN entries right away, rendering two distinct lockout mechanisms ineffective: a Windows-induced 1, 2, or 5-minute timeout, as well as a TPM-induced timeout (2 hours on this device once Hello PIN setup completes, \ref{sec:bruteforce_lockout_conf}).
To show that our attack works, we recorded a demo video of an unoptimized end-to-end attack against the AMD-based OmniBook with a TPM-protected Windows Hello PIN.
The video shows our attacker hardware automatically trying all PINs starting from 1000 until it reaches 1235 (the correct PIN), which unlocks the system.
While the attack ran from 18:27--19:46 (visible on a clock in the timelapse), 19:13--19:35 was spent unsticking our rudimentary brute-forcing script; net time to find the correct PIN was 57 minutes across 236 tried PINs, i.e. \textasciitilde41 PINs per 10 minutes (4.14 tries/minute) -- compared to the TPM lockout's 1 try per 10 minutes.\footnote{Hello PoC video: \url{https://youtu.be/Yq2Wa-nj68g}}
However, the performance of our rudimentary setup -- shared by both attacks -- has large room for optimization, e.g., by restoring the target's original SPI clock speed or writing NV state to flash faster.
\autoref{tab:bruteforce-speeds} calculates worst-case brute-force times directly from this measured rate.
Compared to Microsoft's theoretical 10-minute timeouts per try, this yields a \textasciitilde41x speedup.
Compared to the actual timeout of 2 hours, the speedup is a factor of \textasciitilde497.

\subsection{Generalizing to BitLocker's TPM+PIN Protector}
\label{sec:bitlocker-bruteforce}

The rollback primitive from \autoref{sec:nv-state-rollback-attack} is not specific to Windows Hello: it affects any TPM-sealed secret gated by the Dictionary Attack Protection counters, including BitLocker's \emph{TPM+PIN} protector (\ref{sec:bg_disk_encryption}).
Unlike Hello, BitLocker's pre-boot PIN prompt imposes no additional Windows-side try or timeout limit on top of the TPM's own 32-try budget, so a rollback between boots grants the full 32 tries every cycle.
Using the Winbond leaked-key rollback path (\autoref{sec:nv_rollback_leaked_key}) on the Framework Laptop 13, we demonstrated this live: our interposer rolls back the NV state between reboots, granting the full 32-try TPM budget every boot cycle, and our automation enters PIN attempts at the pre-boot prompt until the correct PIN is found.
Our demo video\footnote{BitLocker PoC video: \url{https://youtu.be/MnplktYLMJw}} shows our attacker hardware trying all PINs from 123222 to 123457 (the correct PIN) in 8 minutes 40 seconds across 236 tries.
After a simple optimization of our KVM's attack loop, the same 236-try campaign took 6 minutes 55 seconds, i.e. \textasciitilde341 tries per 10 minutes (34.1 tries/minute) -- compared to the TPM lockout's 1 try per 10 minutes -- yielding a worst-case time of 20 days for a default-length 6-digit PIN, as shown alongside Hello's numbers in \autoref{tab:bruteforce-speeds}.
Compared to Microsoft's theoretical 10-minute timeout per try, this yields a \textasciitilde341x speedup; compared to the 2-hour timeout we actually observed on this same Framework Laptop (\autoref{sec:bruteforce_lockout_conf}), the speedup is a factor of \textasciitilde4092.

\subsection{Brute-Forcing with NV Rollbacks and TPM Object}
\label{sec:bruteforce_tpm_object}

Finally, we would like to highlight one more attack scenario that deserves attention that we have not, however, experimentally verified at the time of writing:
Recall that at the core, Windows Hello protects a private user key by sealing it with the TPM under a PIN-protected auth policy (\ref{sec:windows_hello}, \ref{sec:bg_tpm}).
Windows stores the TPM object (i.e., the TPM-encrypted user private key) on a disk that, by default, is encrypted on Windows 11.
If, however, an attacker obtains access to that TPM object, for example, because the attacker has temporary access to the \emph{unlocked} device, including access to Windows system files, or the device does not employ disk encryption, brute-force times are even lower.
The attacker may now boot the target device into an attacker-controlled operating system.
By then issuing the same TPM commands (e.g., using \texttt{tpm2-tools}\footnote{\url{https://tpm2-tools.readthedocs.io/en/latest/}}) as Windows does, including the entry of a candidate PIN and a subsequent call to \emph{Unseal}, an attacker gets the ability to check 32 PINs before needing to reboot and rollback the NV state.
This should elevate the attack speed against Windows Hello roughly to the BitLocker's attack speed.
In addition, it could reenable a Windows Hello migration attack as introduced in \cite{kim_security_2021}.

\section{Discussion}
\label{sec:discussion}

Our two end-to-end attacks against Windows Hello and BitLocker (\autoref{sec:hello_bruteforce}) were made possible by two independent attack paths our RPMC research surfaced for rolling back the NV state (including the fTPM state) between reboots, both confirmed end-to-end against a live target:
\begin{enumerate}
    \item Systems relying on Winbond flash chips for RPMC face the risk of an attacker extracting HMAC keys through voltage glitching (\ref{sec:hmac_key_leak}) and interposing SPI traffic to forge counter requests (\ref{sec:nv_rollback_leaked_key}).
    \item AMD-based systems relying on RPMC can be targeted by an interposer without knowledge of the HMAC or Root keys: Targeted suppression of the \rpmcinc command suffices (\ref{sec:nv_rollback_increment_suppression}).
\end{enumerate}
On AMD systems without RPMC flash, neither is needed: reprogramming an old flash image with an SPI programmer suffices (\autoref{sec:nv_state_rollback_protection}).

\paragraph{Mitigations}
While the former issue requires changes to Winbond hardware to fix it at the root (\ref{sec:winbond-mitigations}), the latter issue can, as far as our assessment goes, be addressed by AMD through firmware updates (\ref{sec:amd_mitigations}).

\paragraph{Affected Systems}
To our knowledge, all RPMC-based systems using Winbond flash are vulnerable to interposer attacks using a leaked key (\ref{sec:nv_rollback_leaked_key}).
Furthermore, unless AMD provides firmware fixes\footnote{It is important that these PSP firmware updates prevent rollback to older, vulnerable firmware versions using eFuses or similar means.  Otherwise, an attacker could downgrade the firmware and mount the attack nonetheless.}, all RPMC-based systems with AMD SoCs should be vulnerable to the interposer attacks suppressing increments during NV state operations (\ref{sec:nv_rollback_increment_suppression}).
Moreover, AMD-based systems without RPMC flash appear to lack NV rollback protection altogether: on both such systems we tested, reprogramming the SPI flash with an old image sufficed to roll back the fTPM state (\autoref{sec:nv_state_rollback_protection}).

On AMD systems, software can identify whether RPMC is in use:
Since Linux~5.19, the \code{ccp} driver exports the PSP's security attributes via sysfs, including whether an RPMC-capable SPI flash is present (\code{rpmc\_spirom\_available}) and whether RPMC has been enabled (\code{rpmc\_production\_enabled}) \cite{linux_sysfs_driver_ccp}.\footnote{The PSP exports these attributes only on select SoCs; their absence does not imply the absence of RPMC \cite{fwupd_hsi}.}
The firmware update daemon fwupd summarizes both in its \emph{AMD SPI Replay protections} attribute, displayed by \code{fwupdmgr security} \cite{fwupd_hsi}.
These reports match our results: the Lenovo IdeaPad, whose Authentication Flash supports RPMC commands, reports both attributes as~1 and detected naive rollbacks, whereas the Zen~2 desktop reports both as~0 and accepted them.
Thus, on AMD, either outcome exposes the system to one of the rollback paths above: the \emph{Unverified Increment} with RPMC, or a naive rollback without it.
For Intel systems, we are not aware of a comparable interface; whether they rely on RPMC can, to our knowledge, only be identified through inspection of the mainboard and SPI bus traffic:
If the mainboard hosts an RPMC-capable flash chip (e.g., Winbond's ``W25R'' series), the system likely relies on RPMC.
To confirm this, SPI traffic during boot can be analyzed using a logic analyzer.

When picking a system, it is hard to know beforehand which flash the OEM uses or whether RPMC is used, unless it is openly documented, as in the Framework laptop \cite{framework_intel_schematic}.
In our research, we have not conducted tests on non-x86 laptops, such as ARM-based laptops.
For Qualcomm-based platforms with RPMB-capable storage -- a related concept to RPMC specified in JEDEC's eMMC specification \cite{jesd84} -- Windows 11's Hardware Compatibility Program dictates rollback protection, too (\ref{sec:windows_hcp_rpmc}).

\paragraph{Responsible Disclosures}
We initially informed Winbond about the susceptibility of their chips to glitching in January 2025; Winbond acknowledged the report but did not follow up.
In April 2026, we sent a second report detailing that voltage glitching also leaks the HMAC key itself, together with the resulting impact on Windows Hello and BitLocker, and sent a further reminder in September 2026.
Winbond has since responded that their engineering team is actively evaluating the issue, possibly including a silicon revision, but has not committed to a concrete fix or timeline.

We informed AMD about the increment suppression attack and how to fix it in February 2026 and have followed up, demonstrating the end-to-end attack against Windows Hello in April 2026 and proposing two concrete firmware-fix options together with the BitLocker impact in September 2026.
AMD acknowledged the report and forwarded it for internal review, but has not yet given a technical assessment or committed to a fix.
In November 2026, we informed AMD that naive rollbacks succeed on its platforms without RPMC flash (\autoref{sec:nv_state_rollback_protection}).

Additionally, we reported our end-to-end attack and the enabling vulnerabilities by both AMD and Winbond to Microsoft in April 2026.
Despite running Microsoft-authored code from a dedicated core, Microsoft Pluton's fTPM is similarly affected by AMD's \emph{Unverified Increment} (\ref{sec:unverified_increment}) as AMD's own fTPM is (\ref{sec:interposer_results}).
We also demonstrated our end-to-end attack against Windows Hello to them.
Microsoft closed this case as ``moderate severity'', \emph{defense-in-depth}, below the bar for immediate servicing -- issuing no CVE or bounty, but committing to raise the recommended minimum Hello PIN length to 6 digits.
In September 2026, we further disclosed that the same rollback also defeats BitLocker's TPM+PIN protector, which is brute-forceable even faster than Windows Hello (\ref{sec:bitlocker-bruteforce}); at Microsoft's request, we filed this as a separate report, since it is handled by a different product team and falls outside the scope of their in-progress Hello fix, which predates our BitLocker disclosure.
Microsoft ultimately permitted us to publish the specific Hello mitigation language proposed below (see \emph{Interim Mitigations}), but declined to provide any further written statement for this paper; they expect to issue their own communications once their fix ships, alongside our publication on November 15.

Regarding our minor findings, Intel initially closed our report on the \emph{HMAC Key Reuse} as ``Informative/Rejected'', stating that the responsible business unit does not consider the issue independently exploitable.
In September 2026, Intel reopened the case; Intel PSIRT considers the finding a security weakness rather than a vulnerability, as it only reduces the attacker's effort once a Winbond \emph{HMAC Key Leak} has occurred.
We agree with this classification and adopt it in this version.
Nonetheless, the weakness still upgrades a one-time Winbond \emph{HMAC Key Leak} into a permanent compromise, and Intel could mitigate it by falling back to its pre-RPMC rollback protection on systems with Winbond flash.

JEDEC has responded to the \emph{Integer Overflow}, saying that our suggestions would be forwarded to the formulating committee.

We disclosed the \emph{Integer Overflow} finding to Macronix in April 2025; they confirmed the counter wraparound but, calculating approximately 680 years to reach it at the minimum update interval, closed the case as impractical to exploit.

We disclosed our findings on \emph{Counter Corruption} under glitching to GigaDevice in April 2026.
GigaDevice has engaged constructively, requesting and receiving our fault-injection parameters in September 2026 to attempt reproduction; a fix or formal statement is still pending.

\paragraph{Interim Mitigations}
Since none of the major firmware (AMD) or hardware (Winbond) vulnerabilities have been fixed at the time of writing, Microsoft could still meaningfully raise the bar for our attacks purely through policy, without waiting on vendor firmware or hardware fixes.
Raising the default minimum PIN length by just two digits -- from 4 to 6 for Windows Hello, and from 6 to 8 for BitLocker's TPM+PIN protector -- substantially narrows our NV Rollback attack's window of practicality (\autoref{tab:bruteforce-speeds}): the worst-case retrieval time increases from 40 hours to 168 days for Hello, and from 20 days to 6 years for BitLocker.
This would not eliminate the underlying rollback primitive, but it substantially narrows the window in which our attack remains practical.

\paragraph{RPMC Integration}
Regarding the overall integration trends for RPMC, we are not aware of whether OEMs (e.g., Lenovo) or SoC vendors (e.g., Intel) decide on the rollback prevention method (RPMC or non-RPMC methods).
Microsoft, for its part, does not mandate the method either -- at least not anymore (\ref{sec:windows_hcp_rpmc}).
Our anecdotal evidence on the presence or absence of RPMC in laptops, shown in \autoref{tab:evaluated_systems}, indicates that both AMD and Intel integrated RPMC(-like) flash into systems as early as 2021 and that AMD has at least one Copilot+ certified laptop with RPMC. Intel builds one without RPMC, even though Intel was the initial driver behind the extensions (\ref{sec:rpmc_chronicle}).

\paragraph{RPMC Security}

We will now debate the overall benefit of RPMC integration compared to pre-RPMC rollback prevention of the platform NV state.
It can be argued that RPMC logically places the rollback counters where the protected NV state data itself lives: SPI flash.
However, this comes at several disadvantages:
First and foremost, it reintroduces an external bus that fTPMs were meant to eliminate, exposing security operations to it again.
This requires cryptographic legwork:
The SoC and flash chip must generate and exchange the shared \emph{Root Key} in a trusted OEM environment.
Then, based on random data, this key must be rotated to allow signature creation and verification on both ends.
While coprocessors like the AMD PSP and Intel CSME clearly already have the capabilities for the \emph{Root Key} storage as well as the cryptographic operations, although not openly documented nor accessible to 3rd parties, our research shows that flash vendors fail to implement these mechanisms securely on their end.
The fault injection attacks used in this research do not exhaust the full toolset of hardware attacks; other methods, such as electromagnetic or laser fault injection, as well as side-channel analysis, come to mind.
Regarding the integrity of the data that the SPI NOR flash delivers, unreliable reads can be detected by parity or signature checks, as widely employed in the industry. 
With its additional RPMC responsibilities, the SPI flash suddenly becomes the Achilles heel of NV state integrity and, as a consequence, the fTPM's integrity.
On Windows Hello for Business, this ends up protecting enterprise SSO (\ref{sec:whfb_sso}).

Compare this to pre-RPMC rollback protection of NV state: Our experiments indicated a functional, self-contained rollback protection implemented by Intel that we were unable to circumvent naively, presumably the ME's timer-based counter (\autoref{sec:nv_state_rollback_protection}).\footnote{This is, unless the fTPM-hosting TEE itself is targeted as with the faulTPM attack\cite{faultpm_2023}.}
It might still exist and be ready to activate even on systems shipped with an RPMC configuration.
\textbf{Based on our current understanding, we believe that Intel should switch back from RPMC-backed to pre-RPMC rollback prevention through firmware updates to reinstantiate fTPM security even on systems with insecure RPMC flash.}
For AMD, the situation is reversed: our experiments indicate no such fallback (\autoref{sec:nv_state_rollback_protection}), so RPMC is currently AMD's only means of NV rollback protection.
On AMD platforms without RPMC flash, an SPI programmer suffices to roll back the fTPM state, which arguably falls short of the rollback protection that Microsoft's hardware compatibility programs require \cite{win10_hcp, win11_hcp}.
Rather than abandoning RPMC, AMD should verify its increments (\autoref{sec:unverified_increment}) and provide rollback protection for platforms without RPMC flash.

\paragraph{RPMC Repairability}

Beyond these security concerns, RPMC integration also affects repairability. Anecdotally, during our research, the Framework's RPMC flash failed due to a faulty supply voltage.
Normally, we would have been able to desolder the defective flash and replace it with a new one.
However, the \emph{Root Key} of the broken chip is unknown to us, and the Intel SoC is in a post-provisioning state that does not accept other \emph{Root Keys}. 
As a result, our system is effectively bricked.
When we contacted Framework, a company known for its high repairability standards, about this, they said that the repair process for this board only allows a full board replacement.
We cannot say whether this is due to missing microsoldering capabilities at the repair centers or to a missing RPMC reprovisioning flow in Framework's or Intel's firmware.
Nonetheless, this part pairing would not exist without RPMC.

\section{Conclusion}
\label{sec:conclusion}

The shift from discrete to firmware TPMs brought a meaningful security benefit: by running TPM logic inside the SoC's security coprocessor, the exposed SPI bus -- previously exploitable for passive and active attacks -- is no longer on the critical path for TPM communication.
RPMC-backed rollback protection partially undoes that progress by re-externalizing a security-critical part of the fTPM state -- its anti-hammering counters -- to the very bus that fTPMs had incidentally rendered inaccessible.

Our first security evaluation of RPMC deployments in shipping x86 systems reveals multiple independent implementation flaws that compromise this protection.
Winbond -- the world's leading NOR flash vendor -- leaks its HMAC authentication key under voltage glitching, enabling arbitrary counter forgery in under a second.
Notably, Winbond's own datasheet reproduces JEDEC's power-glitch-aware requirement almost verbatim, yet our results show a simple crowbar glitch defeats it.
Intel's NV controller amplifies this by reusing a constant session key across all boots, reducing the full key-extraction attack to a single, one-time glitch.
AMD's NV controller, independently of any flash vulnerability, never verifies that an issued increment was actually applied, allowing a low-cost interposer to silently suppress all counter increments and thus permit unlimited NV state rollbacks.

We demonstrated the real-world consequences of both flaws with end-to-end brute-force attacks: AMD's \emph{Unverified Increment} against the Windows Hello PIN, and Winbond's \emph{HMAC Key Leak} against BitLocker's TPM+PIN protector, both on locked, encrypted hardware; the former even Copilot+-certified.
By combining fTPM state rollbacks with automated PIN entry, we achieve approximately 41 times the attempt rate of the default TPM lockout regime, reducing the expected time to crack a 4-digit numeric PIN on a locked device from 69 days to 40 hours.
The impact is not limited to lockscreen bypass: a brute-forced Hello PIN can unlock device-bound keys, enable Hello-based enterprise SSO, and, in some configurations, allow an attacker to satisfy an otherwise multi-factor-protected authentication.
Lastly, the same rollback primitive generalizes beyond Hello: we also demonstrate it live against BitLocker's own TPM+PIN pre-boot protector, which grants the full 32-try TPM budget on every boot without any Windows-imposed throttling, reducing the expected time to crack a default-length 6-digit PIN on a locked device from 19 years (with the laxer 10-minute TPM lockout) to 20 days -- a \textasciitilde341 times speedup.
On AMD systems without RPMC-capable flash, an attacker may not even need either flaw: on Zen~2 and Zen~5 desktops, naive rollbacks of the NV state with an SPI programmer succeeded.

Our results show that the current RPMC ecosystem -- spanning specification gaps, flash vendor implementations, and SoC controller design -- is not up to its task.
We recommend that AMD deploys a firmware fix to verify increment success and that Winbond addresses the key-leak vulnerability in hardware, or, alternatively, that Intel considers reverting to pre-RPMC rollback protection on affected systems as a short-term measure.
We have responsibly disclosed all findings to Winbond, Macronix, GigaDevice, Intel, AMD, JEDEC, and Microsoft, and we make our tools and interposer design publicly available to facilitate further research and vendor remediation.

\begin{acks}
This research was funded by the German Federal Ministry of Research, Technology and Space within the DI-SIGN-HEP Project Grant 16KIS2074.
\end{acks}

\section*{Open Science}
\label{sec:open-science}
The artifacts of our paper are published as a Git repository\footnote{\url{https://github.com/mattifinder/rpmc_attacks}} that contains the interposer firmware (both board revisions), our interposer PCB designs, scripts used to mount the KVM attack, build and run descriptions, the leaked HMAC keys of our test system's Winbond flash chip, and an excerpt of the traces we used in finding our attack -- per platform, one boot trace and a trace showing the behavior when entering an incorrect Windows Hello PIN.

We also provide two short timelapse PoC videos on YouTube where we brute-forced 236 PINs: one of the Windows Hello attack\footnote{\url{https://youtu.be/Yq2Wa-nj68g}} on the AMD laptop taking a net 57 minutes and one of the BitLocker attack\footnote{\url{https://youtu.be/MnplktYLMJw}} on the Intel laptop taking 8 minutes 40 seconds (6 minutes 55 seconds after optimizing our KVM's attack loop, see \autoref{sec:bitlocker-bruteforce}).

\section*{Changes from the CCS '26 Version}\phantomsection
\label{sec:changelog}

This version differs from the paper published at ACM CCS~'26 as follows:

\begin{itemize}
  \item \emph{AMD rollback protection without RPMC.}
    In the CCS version, we classified the Lenovo IdeaPad 5 Pro 16ACH16 as a system without RPMC.
    After the final deadline, we found that its Winbond W74M flash is a so-called \emph{Authentication Flash}, which is functionally identical to RPMC flash and uses the same commands; SPI traces show that the system uses them during boot.
    Furthermore, we evaluated three additional desktop systems without RPMC flash.
    On the two AMD-based ones (Zen~2 and Zen~5), naive rollbacks of the NV state succeeded and reset the fTPM's dictionary attack counters.
    Contrary to our original assumption, AMD SoCs therefore likely do not feature standalone rollback protection for their NV state.
    The Intel-based one (Core i5-14400) detected the rollback, consistent with the ME's timer-based anti-replay counter described by Sklyarov \cite{sklyarov_mfs_explained}.
    We updated the abstract, \autoref{tab:evaluated_systems}, the contributions in \autoref{sec:introduction}, \autoref{sec:nv_state_rollback_protection}, the \emph{RPMC Integration} and \emph{RPMC Security} paragraphs in \autoref{sec:discussion}, and \autoref{sec:conclusion} accordingly, and we limited our recommendation to fall back to pre-RPMC rollback protection to Intel.
    In \emph{Affected Systems} (\autoref{sec:discussion}), we added that AMD systems without RPMC are affected as well and described how Linux and fwupd report whether an AMD system relies on RPMC.
  \item \emph{BitLocker brute-force speed.}
    An optimization of our KVM's attack loop reduced the duration of the 236-try BitLocker campaign from 8 minutes 40 seconds to 6 minutes 55 seconds (34.1 instead of 27.2 tries/minute).
    The worst-case time for a default-length 6-digit BitLocker PIN thus drops from 26 to 20 days.
    We updated the abstract, \autoref{sec:introduction}, \autoref{tab:bruteforce-speeds}, \autoref{sec:bitlocker-bruteforce}, \autoref{sec:discussion}, and \autoref{sec:conclusion} accordingly.
  \item \emph{Disclosure updates.}
    We added that we informed AMD about the naive rollbacks, and, following Intel PSIRT's assessment, we now classify the \emph{HMAC Key Reuse} as a security weakness rather than a vulnerability (\autoref{sec:hmac_key_reuse}, \autoref{sec:discussion}).
  \item \emph{Shortened abstract.}
    We shortened the abstract to fit arXiv's length limit; the findings it summarizes are unchanged.
\end{itemize}

\newpage

\renewcommand{\bibfont}{\bibliofont\linespread{0.96}\selectfont}
\bibliographystyle{ACM-Reference-Format}
\bibliography{refs}

\appendix

\end{document}